\documentclass[12pt]{article}

\usepackage{amsmath}
\usepackage{amssymb}
\usepackage{colortbl}

\allowdisplaybreaks

\begin{document}

\baselineskip=18.8pt plus 0.2pt minus 0.1pt

\makeatletter

\@addtoreset{equation}{section}
\renewcommand{\theequation}{\thesection.\arabic{equation}}
\renewcommand{\thefootnote}{\fnsymbol{footnote}}
\newcommand{\beq}{\begin{equation}}
\newcommand{\eeq}{\end{equation}}
\newcommand{\bea}{\begin{eqnarray}}
\newcommand{\eea}{\end{eqnarray}}
\newcommand{\nn}{\nonumber\\}
\newcommand{\hs}[1]{\hspace{#1}}
\newcommand{\vs}[1]{\vspace{#1}}
\newcommand{\p}{\partial}
\newcommand{\bra}[1]{\left\langle  #1 \right\vert }
\newcommand{\ket}[1]{\left\vert #1 \right\rangle }
\newcommand{\vev}[1]{\left\langle  #1 \right\rangle }

\makeatother

\begin{titlepage}
\title{
\hfill\parbox{4cm}
{\normalsize\tt 
}\\
\vspace{1cm}
Counting SO(9)$\times$SU(2) representations in coordinate independent state space of SU(2) Matrix Theory
}
\author{Yoji Michishita
\thanks{
{\tt michishita@edu.kagoshima-u.ac.jp}
}
\\[7pt]
{\it Department of Physics, Faculty of Education, Kagoshima University}\\
{\it Kagoshima, 890-0065, Japan}
}

\date{\normalsize September, 2010}
\maketitle

\begin{abstract}
\normalsize
We consider decomposition of coordinate independent states into SO(9)$\times$SU(2) representations
in SU(2) Matrix theory. To see what and how many representations appear in the decomposition,
we compute the character, which is given by a trace over the coordinate independent states,
and decompose it into the sum of products of SO(9) and SU(2) characters.
\end{abstract}
\end{titlepage}

\clearpage
\section{Introduction}

Matrix theory, which is expected to be a correct description of M-theory, is a quantum mechanics
with two sets of operators: bosonic coordinate matrices $X_i^a$ and fermionic matrices $\theta_\alpha^a$.
This quantum mechanics has SO(9) symmetry of space rotation, and gauge symmetry SU($N$).
To investigate the structure of wavefunctions in this theory, it is necessary to know the structure of
the space spanned by coordinate independent states i.e. states constructed only by $\theta_\alpha^a$.
Especially we want to know what representations of SO(9) and SU($N$) those states form.
However the number of states are enormous even if we only take coordinate independent states,
and it makes explicit construction of representations difficult. 

In this paper we count numbers of representations appearing in the space of coordinate independent states
in the case of SU(2) gauge group, avoiding explicit construction of representations.
To do it efficiently we employ the notion of characters in group theory:
We introduce $\chi$, a trace of a group element of SO(9)$\times$SU(2) over the coordinate independent states.
If we take an appropriate basis of the states, $\chi$ can be calculated explicitly,
and by decomposing it into sum of products of SO(9) and SU(2) characters, we can immediately read off
what and how many representations appear in the space of coordinate independent states.
Similar analyses have been made in \cite{hlt08} for SO(7)$\times$SU(2) singlets,
and in \cite{trze07} for SU($N$) gauge group singlets.
As a byproduct of our calculation we can give another proof of the uniqueness of SO(9)$\times$SU(2) singlet
proven in \cite{hlt08,ht10}.

In the next section we compute $\chi$ and perform the decomposition. In Appendix A and B
we collect information on group theory necessary for the analysis.
Calculations are made with the help of symbolic manipulation program Mathematica.
\section{SO(9)$\times$SU(2) character}

Matrix theory has real Grassmann odd operators $\theta_\alpha^a$,
where $\alpha=1,2,\dots,16$ is an SO(9) spinor index and $a=1,2,\dots,N^2-1$ is an adjoint index
of the gauge group SU($N$). Their anticommutation relation is
\beq
\{\theta_\alpha^a,\theta_\beta^b\}=\delta_{\alpha\beta}\delta^{ab}.
\eeq
For SU(2) gauge group, we have $16\times 3=48$ operators, and half of those are regarded as creation operators
and the other half as annihilation operators. Then we can construct $2^{48/2}=2^{24}$ states.
If we fix the adjoint index $a$ then the 8 creation operators give 256 states, which are classified
into 44-dimensional symmetric traceless representation, 84-dimensional 3-rank antisymmetric representation, and
128-dimensional vector-spinor representation of SO(9). If we take the adjoint index into account, the decomposition
of $2^{24}$ states into SO(9)$\times$SU(2) representations is not immediately clear.
To construct gauge invariant wavefunctions it is imoportant to know it.
To this end, we introduce the character $\chi$, which is given by the trace over the $2^{24}$-dimensional space
and is a function of parameters $x_1,x_2,x_3,x_4$ and $y$:
\beq
\chi=\text{tr}\big[\exp(ix_1J_{12}+ix_2J_{34}+ix_3J_{56}+ix_4J_{78}+iyg^1)\big],
\eeq
where $J_{ij}=-\frac{i}{4}\theta^a_\alpha(\gamma_{ij})_{\alpha\beta}\theta^a_\beta$ are SO(9) generators and
$g^a=\frac{i}{2}\epsilon_{abc}\theta^b_\alpha\theta^c_\alpha$ are SU(2) generators.
In addition we define $\Tilde{\chi}$ by the following, with fermion number operator insertion $(-1)^F$:
\beq
\Tilde{\chi}=\text{tr}\big[(-1)^F\exp(ix_1J_{12}+ix_2J_{34}+ix_3J_{56}+ix_4J_{78}+iyg^1)\big].
\eeq
Here we define traces of states as the sum of contributions from boson states and fermion states.
Therefore $\Tilde{\chi}$ gives the difference of contributions from boson and fermion states.

Since the trace does not depend on choice of orthogonal basis of states, we will take
one which makes calculation of the characters easier. 
Although $\theta_\alpha^a$ is real, let us temporarily consider the case where
$\theta_\alpha^a$ is complex and their anticommutation relation is
$\{\theta_\alpha^a,(\theta_\beta^b)^\dagger\}=\delta_{\alpha\beta}\delta^{ab}$. 
(In other words, $\theta_\alpha^a$ is given by two copies of the real one $\theta^{(1)}$ and $\theta^{(2)}$:
 $\theta_\alpha^a=(\theta^{(1)a}_\alpha+i\theta^{(2)a}_\alpha)/\sqrt{2}$.) Then
we could regard $(\theta_\alpha^a)^\dagger$ as creation operators. Since these operators are covariant under
both of SO(9) and SU(2) transformations, the character were given just by sum of characters of antisymmetric
tensor product representations Alt$_n$[(SO(9) spinor)$\times$(SU(2) adjoint)],
which can be calculated by Frobenius formula (See Appendix \ref{appa}).
(Such calculation has been done in \cite{trze07}.)
However in our case where $\theta_\alpha^a$ are real,
creation and annihilation operators cannot be separated without losing manifest covariance.
Therefore we will take a different way and calculation will be harder than the complex case.
First we define $\theta_\alpha^\pm$ as
\bea
\theta_\alpha^\pm=\frac{1}{\sqrt{2}}(\theta_\alpha^2\pm i\theta_\alpha^3).
\eea
Note that $(\theta_\alpha^\pm)^\dagger=\theta_\alpha^\mp$.
Then nontrivial anticommutation relations are given by
\beq
\{\theta_\alpha^1,\theta_\beta^1\}=\delta_{\alpha\beta},\quad 
\{\theta_\alpha^-,\theta_\beta^+\}=\delta_{\alpha\beta}.
\eeq
A vacuum $\ket{0}$ for these operators is defined as follows:
\beq
\theta_\alpha^-\ket{0}=0,
\eeq
and $\theta_\alpha^+$ work as creation operators on this vacuum.
Since it is not necessary in the following, we do not specify the action of $\theta^1_\alpha$
on $\ket{0}$. Of course we can make different choices of vacuum and creation operators. All of them 
give the same coordinate independent state space, and make part of SO(9)$\times$SU(2) symmetry not manifest.
As a result those vacua cannot be SO(9)$\times$SU(2) singlet.
Our choice makes SO(9) symmetry manifest at the expense of SU(2) symmetry, and
each step of computation of the characters in the following has manifest SO(9) symmetry.
At the end of the computation we will recover manifest SU(2) symmetry.
\footnote{In \cite{hlt08} the coordinate independent state space for fixed gauge indices is constructed, and
its natural extension to the case with gauge indices is given by
$\lambda_\alpha^a\equiv \frac{1}{\sqrt{2}}(\theta_\alpha^a+i\theta_{\alpha+8}^a)~(\alpha=1,\dots,8),
~(\lambda_\alpha^a)^\dagger\ket{0'}=0$.
This retains manifest gauge 
symmetry at the expense of SO(9) symmetry. We retain SO(9) instead of SU(2) because the structure of 
characters of SU(2) is simpler than that of SO(9) and it is easier to recover SU(2) than SO(9).
The vacuum $\ket{0'}$ is related to our $\ket{0}$ by $\ket{0}=\prod_{\alpha=1}^8\lambda_\alpha^-\ket{0'}$.
}

We can immediately see that actions of $\theta_\alpha^1$ and $\theta_\alpha^\pm$ can be considered separately,
and the characters are decomposed into two parts corresponding to them: 
\beq
\chi=\chi_{\theta^1}\chi_{\theta^\pm},\quad \Tilde{\chi}=\Tilde{\chi}_{\theta^1}\Tilde{\chi}_{\theta^\pm}.
\eeq
Since $[g^1, \theta_\alpha^1]=0$,
$\chi_{\theta^1}$ and $\Tilde{\chi}_{\theta^1}$ can be readily computed.
Indeed $\chi_{\theta^1}$ is just the sum of SO(9) characters of 
2-rank symmetric traceless, 3-rank antisymmetric, and vector-spinor representation:
\beq
\chi_{\theta^1}=\chi_{[2000]}+\chi_{[0010]}+\chi_{[1001]},\quad
\Tilde{\chi}_{\theta^1}=\chi_{[2000]}+\chi_{[0010]}-\chi_{[1001]},
\eeq
where SO(9) representations are indicated by Dynkin labels $[q_1q_2q_3q_4]$. 
See Appendix \ref{appb} for more infomation and notation for SO(9) characters.

Next we compute $\chi_{\theta^\pm}$ and $\Tilde{\chi}_{\theta^\pm}$.
States are classified by the number of $\theta_\alpha^+$ on $\ket{0}$:
\beq
\ket{0},\quad \theta_\alpha^+\ket{0},\quad \theta_{\alpha_1}^+\theta_{\alpha_2}^+\ket{0},\quad
\dots,\quad\theta_{\alpha_1}^+\dots\theta_{\alpha_{16}}^+\ket{0}.
\eeq
Since $g^1=8-\theta_\alpha^+\theta_\alpha^-$
and $\theta_\alpha^+\theta_\alpha^-$ works as the number operator for $\theta_\alpha^+$,
in the sector of $n$ $\theta_\alpha^+$, the factor $\exp(iyg^1)$ in the characters
gives $e^{i(8-n)y}$. Obviously this sector forms $n$-rank antisymmetric 
product representation of SO(9) spinor, and therefore contribution to $\chi_{\theta^\pm}$ from this sector is
given by
\beq
e^{i(8-n)y}~\chi(\text{Alt}_n(\text{spinor})).
\eeq 
$\chi(\text{Alt}_n(\text{spinor}))$ can be calculated by Frobenius formula.
In fact, sectors of $n$ $\theta_\alpha^+$ and of $16-n$ $\theta_\alpha^+$ are in the
same representation of SO(9), because states $\theta_{\alpha_1}^+\dots\theta_{\alpha_n}^+\ket{0}$
are also expressed as $\epsilon_{\alpha_1\dots\alpha_n\alpha_{n+1}\dots\alpha_{16}}
\theta_{\alpha_1}^+\dots\theta_{\alpha_n}^+\ket{0}$.
Indeed straightforward calculation shows $\chi(\text{Alt}_{16-n}(\text{spinor}))=\chi(\text{Alt}_n(\text{spinor}))$.

Then the total contributions are
\bea
\chi_{\theta^\pm} & = &
\sum_{n=0}^7~[e^{i(8-n)y}+e^{-i(8-n)y}]~\chi(\text{Alt}_n(\text{spinor}))
+\chi(\text{Alt}_8(\text{spinor})),
\label{tchar0-1}
\\
\Tilde{\chi}_{\theta^\pm} & = &
\sum_{n=0}^7~[e^{i(8-n)y}+e^{-i(8-n)y}]~(-1)^n\chi(\text{Alt}_n(\text{spinor}))
+\chi(\text{Alt}_8(\text{spinor})).
\label{tchar0-2}
\eea
As is well-known, representations of SU(2) are labeled by nonnegative half integers (spins), and for
spin $n$ representation eigenvalues of $g^1$ are $-n,-n+1,\dots,n-1,n$ and the character $\chi^{\text{SU(2)}}_n$
for this representation is given by
\beq
\chi^{\text{SU(2)}}_n=\text{tr}_{\text{spin}\,n}e^{iyg^1}
=e^{i(-n)y}+e^{i(-n+1)y}+\dots+e^{i(n-1)y}+e^{iny}.
\eeq
Note that $e^{iny}+e^{-iny}=\chi^{\text{SU(2)}}_n-\chi^{\text{SU(2)}}_{n-1}$. Using this we can rewrite
$y$ dependent part of \eqref{tchar0-1} and \eqref{tchar0-2}. Then the total characters are given by the following,
in the forms which make decomposition into SU(2) representations manifest: 
\beq
\chi=\sum_{n=0}^8\chi^{\text{SU(2)}}_n\chi^{\text{SO(9)}}_n,\quad
\Tilde{\chi}=\sum_{n=0}^8\chi^{\text{SU(2)}}_n\Tilde{\chi}^{\text{SO(9)}}_n,
\eeq
where
\bea
\chi^{\text{SO(9)}}_n & = & \begin{cases} 
\chi_{\theta^1}[\chi(\text{Alt}_{8-n}(\text{spinor}))-\chi(\text{Alt}_{7-n}(\text{spinor}))]
& (n=0,\dots,7), \\
\chi_{\theta^1} & (n=8), \end{cases}
\label{so9n1}
\\
\Tilde{\chi}^{\text{SO(9)}}_n & = & \begin{cases}
(-1)^n\Tilde{\chi}_{\theta^1}[\chi(\text{Alt}_{8-n}(\text{spinor}))+\chi(\text{Alt}_{7-n}(\text{spinor}))]
& (n=0,\dots,7), \\
\Tilde{\chi}_{\theta^1} & (n=8). \end{cases}
\label{so9n2}
\eea
$\chi^{\text{SO(9)}}_n$ and $\Tilde{\chi}^{\text{SO(9)}}_n$
can be decomposed further, into contributions from boson states and fermion states, denoted by
$\chi^{\text{SO(9),B}}_n$ and $\chi^{\text{SO(9),F}}_n$
respectively:
\beq
\chi^{\text{SO(9),B}}_n=\frac{1}{2}(\chi^{\text{SO(9)}}_n+\Tilde{\chi}^{\text{SO(9)}}_n),\quad
\chi^{\text{SO(9),F}}_n=\frac{1}{2}(\chi^{\text{SO(9)}}_n-\Tilde{\chi}^{\text{SO(9)}}_n).
\eeq
Thus we have obtained explicit expressions for the characters, because
we know explicit expressions of $\chi_{\theta^1}, \Tilde{\chi}_{\theta^1}$ and 
$\chi(\text{Alt}_n(\text{spinor}))$.
However the expressions \eqref{so9n1} and \eqref{so9n2} do not tell us what SO(9) irreducible
representations they contain.
So our next task is to decompose \eqref{so9n1} and \eqref{so9n2} into the sums of SO(9) characters.
Since $\chi^{\text{SO(9),B}}_n$ and $\chi^{\text{SO(9),F}}_n$ are given in the form of products
of SO(9) characters, the decomposition can be performed by decomposing tensor product representations 
of corresponding representations one by one, or by using the following orthogonality relation:
\beq
\frac{1}{2^4\cdot 4!}\prod_{i=1}^4\left(\int_0^{2\pi}\frac{dx_i}{2\pi}\right)[D_{\rho}]^2
\chi_{[q'_1,q'_2,q'_3,q'_4]}\chi_{[q_1,q_2,q_3,q_4]}
=\delta_{q_1',q_1}\delta_{q_2',q_2}\delta_{q_3',q_3}\delta_{q_4',q_4},
\eeq
where $D_\rho$ is defined in Appendix \ref{appb}.
We take the latter method. We just computed integrals of products of SO(9) characters and 
$\chi^{\text{SO(9),B}}_n$ or $\chi^{\text{SO(9),F}}_n$ using Mathematica
and determined the decompositions completely.
Then from them we can immediately read off
what SO(9)$\times$SU(2) representations our $2^{24}$-dimensional space are decomposed into,
and the multiplicities of those representations. The result is given in Table \ref{reptable}.

\begin{table}
\begin{center}
{\tiny
\begin{tabular}{|r|r|c|c|c|c|c|c|c|c|c|} \hline
\multicolumn{2}{|c|}{SO(9)} & \multicolumn{9}{c|}{SU(2) spin} \\ \hline
representation & dimension & 0 & 1 & 2 & 3 & 4 & 5 & 6 & 7 & 8 \\ \hline
{}[0, 0, 0, 0] & 1 		& 1 &   &   &   &   &   & 1 &   &   \\
{}[1, 0, 0, 0] & 9 		&   & 1 &   & 1 &   & 1 &   & 1 &   \\
{}[0, 1, 0, 0] & 36 	&   & 1 & 1 & 1 & 1 & 1 & 1 & 1 &   \\
{}[0, 0, 1, 0] & 84 	& 2 &   & 2 & 1 & 2 & 1 & 2 &   & 1 \\ \rowcolor[gray]{0.9}
{}[0, 0, 0, 1] & 16 	&   & 1 & 1 &   & 1 & 1 & 1 & 1 &   \\
{}[2, 0, 0, 0] & 44 	& 1 &   & 2 &   & 2 &   & 1 &   & 1 \\
{}[0, 2, 0, 0] & 495 	& 1 &   & 2 & 1 & 2 & 1 & 1 &   &   \\
{}[0, 0, 2, 0] & 1980 	& 2 &   & 2 & 1 & 1 &   & 1 &   &   \\
{}[0, 0, 0, 2] & 126 	&   & 2 & 1 & 2 & 1 & 2 & 1 & 1 &   \\
{}[1, 1, 0, 0] & 231 	&   & 2 & 2 & 2 & 2 & 2 & 1 & 1 &   \\
{}[1, 0, 1, 0] & 594 	&   & 3 & 2 & 4 & 2 & 3 & 1 & 1 &   \\ \rowcolor[gray]{0.9}
{}[1, 0, 0, 1] & 128 	& 1 & 2 & 2 & 3 & 2 & 2 & 2 & 1 & 1 \\
{}[0, 1, 1, 0] & 1650 	&   & 2 & 2 & 3 & 2 & 2 & 1 &   &   \\ \rowcolor[gray]{0.9}
{}[0, 1, 0, 1] & 432 	& 1 & 2 & 3 & 3 & 3 & 3 & 2 & 1 &   \\ \rowcolor[gray]{0.9}
{}[0, 0, 1, 1] & 768 	&   & 2 & 3 & 2 & 3 & 2 & 1 & 1 &   \\
{}[3, 0, 0, 0] & 156 	&   & 1 &   & 2 &   & 1 &   &   &   \\
{}[0, 3, 0, 0] & 4004 	&   &   &   & 1 &   &   &   &   &   \\
{}[0, 0, 3, 0] & 23595 	& 1 &   &   &   &   &   &   &   &   \\ \rowcolor[gray]{0.9}
{}[0, 0, 0, 3] & 672 	& 1 & 1 & 1 & 2 & 1 & 1 & 1 &   &   \\
{}[2, 1, 0, 0] & 910 	& 1 & 1 & 2 & 2 & 2 & 1 & 1 &   &   \\
{}[2, 0, 1, 0] & 2457 	& 2 & 1 & 4 & 2 & 3 & 1 & 1 &   &   \\ \rowcolor[gray]{0.9}
{}[2, 0, 0, 1] & 576 	& 1 & 2 & 3 & 3 & 3 & 2 & 1 & 1 &   \\
{}[1, 2, 0, 0] & 2574 	&   & 1 & 1 & 2 & 1 & 1 &   &   &   \\
{}[1, 0, 2, 0] & 12012 	&   & 2 & 1 & 1 &   &   &   &   &   \\
{}[1, 0, 0, 2] & 924 	& 1 & 2 & 4 & 3 & 4 & 2 & 2 & 1 &   \\
{}[0, 2, 1, 0] & 15444 	& 1 &   & 1 &   & 1 &   &   &   &   \\ \rowcolor[gray]{0.9}
{}[0, 2, 0, 1] & 4928 	&   & 1 & 2 & 2 & 2 & 1 &   &   &   \\
{}[0, 1, 2, 0] & 27027 	&   & 1 &   &   &   &   &   &   &   \\
{}[0, 1, 0, 2] & 2772 	& 1 & 2 & 3 & 3 & 3 & 2 & 1 &   &   \\ \rowcolor[gray]{0.9}
{}[0, 0, 2, 1] & 13200 	&   & 1 & 1 &   &   &   &   &   &   \\
{}[0, 0, 1, 2] & 4158 	&   & 2 & 1 & 2 & 1 & 1 &   &   &   \\
{}[1, 1, 1, 0] & 9009 	&   & 2 & 3 & 2 & 2 & 1 &   &   &   \\ \rowcolor[gray]{0.9}
{}[1, 1, 0, 1] & 2560 	& 1 & 3 & 4 & 5 & 4 & 3 & 2 &   &   \\ \rowcolor[gray]{0.9}
{}[1, 0, 1, 1] & 5040 	& 1 & 3 & 4 & 4 & 3 & 2 & 1 &   &   \\ \rowcolor[gray]{0.9}
{}[0, 1, 1, 1] & 12672 	& 1 & 2 & 2 & 2 & 1 & 1 &   &   &   \\
{}[4, 0, 0, 0] & 450 	& 1 &   & 1 &   & 1 &   &   &   &   \\
{}[0, 0, 0, 4] & 2772 	&   &   & 1 &   & 1 &   &   &   &   \\
{}[3, 1, 0, 0] & 2772 	&   & 1 & 1 & 1 & 1 &   &   &   &   \\
{}[3, 0, 1, 0] & 7700 	&   & 2 & 1 & 2 &   &   &   &   &   \\ \rowcolor[gray]{0.9}
{}[3, 0, 0, 1] & 1920 	&   & 2 & 2 & 2 & 2 & 1 &   &   &   \\ \rowcolor[gray]{0.9}
{}[1, 0, 0, 3] & 4608 	& 1 & 1 & 2 & 2 & 2 & 1 &   &   &   \\ \rowcolor[gray]{0.9}
{}[0, 1, 0, 3] & 12672 	&   & 1 & 1 & 1 & 1 &   &   &   &   \\ \rowcolor[gray]{0.9}
{}[0, 0, 1, 3] & 16896 	&   &   &   & 1 &   &   &   &   &   \\
{}[2, 2, 0, 0] & 8748 	& 1 &   & 1 &   & 1 &   &   &   &   \\
{}[2, 0, 2, 0] & 44352 	& 1 &   & 1 &   &   &   &   &   &   \\
{}[2, 0, 0, 2] & 3900 	&   & 3 & 2 & 4 & 2 & 2 &   &   &   \\
{}[0, 2, 0, 2] & 27456 	&   & 1 &   & 1 &   &   &   &   &   \\
{}[2, 1, 1, 0] & 31500 	&   & 1 & 1 & 1 &   &   &   &   &   \\ \rowcolor[gray]{0.9}
{}[2, 1, 0, 1] & 9504 	& 1 & 2 & 3 & 3 & 2 & 1 &   &   &   \\ \rowcolor[gray]{0.9}
{}[2, 0, 1, 1] & 19712 	& 1 & 2 & 3 & 2 & 1 &   &   &   &   \\ \rowcolor[gray]{0.9}
{}[1, 2, 0, 1] & 24192 	&   & 1 & 1 & 1 & 1 &   &   &   &   \\
{}[1, 1, 0, 2] & 15444 	& 1 & 2 & 3 & 3 & 2 & 1 &   &   &   \\ \rowcolor[gray]{0.9}
{}[1, 0, 2, 1] & 76032 	&   & 1 &   &   &   &   &   &   &   \\
{}[1, 0, 1, 2] & 25740 	& 1 & 1 & 2 & 1 & 1 &   &   &   &   \\
{}[0, 1, 1, 2] & 60060 	&   &   & 1 &   &   &   &   &   &   \\ \rowcolor[gray]{0.9}
{}[1, 1, 1, 1] & 65536 	& 1 & 1 & 1 & 1 &   &   &   &   &   \\
{}[5, 0, 0, 0] & 1122 	&   & 1 &   &   &   &   &   &   &   \\
{}[4, 1, 0, 0] & 7140 	&   &   & 1 &   &   &   &   &   &   \\
{}[4, 0, 1, 0] & 20196 	& 1 &   & 1 &   &   &   &   &   &   \\ \rowcolor[gray]{0.9}
{}[4, 0, 0, 1] & 5280 	& 1 & 1 & 1 & 1 &   &   &   &   &   \\
{}[1, 0, 0, 4] & 18018 	&   &   &   & 1 &   &   &   &   &   \\
{}[3, 0, 0, 2] & 12375 	& 1 & 1 & 2 & 1 & 1 &   &   &   &   \\ \rowcolor[gray]{0.9}
{}[2, 0, 0, 3] & 18480 	&   & 1 & 1 & 1 & 1 &   &   &   &   \\ \rowcolor[gray]{0.9}
{}[3, 1, 0, 1] & 27648 	&   & 1 & 1 & 1 &   &   &   &   &   \\ \rowcolor[gray]{0.9}
{}[3, 0, 1, 1] & 59136 	&   & 1 & 1 &   &   &   &   &   &   \\ \rowcolor[gray]{0.9}
{}[1, 1, 0, 3] & 67200 	&   &   & 1 &   &   &   &   &   &   \\
{}[2, 1, 0, 2] & 54675 	&   & 1 & 1 & 1 &   &   &   &   &   \\
{}[2, 0, 1, 2] & 96228 	&   & 1 &   &   &   &   &   &   &   \\
{}[6, 0, 0, 0] & 2508 	& 1 &   &   &   &   &   &   &   &   \\ \rowcolor[gray]{0.9}
{}[5, 0, 0, 1] & 12672 	&   & 1 &   &   &   &   &   &   &   \\
{}[4, 0, 0, 2] & 32725 	&   & 1 &   &   &   &   &   &   &   \\ \rowcolor[gray]{0.9}
{}[3, 0, 0, 3] & 56320 	& 1 &   &   &   &   &   &   &   & \\ \hline
\end{tabular}
}
\end{center}
\caption{Multiplicities of SO(9)$\times$SU(2) representations in the $2^{24}$-dimensional space.
Shaded rows indicate contributions from fermion states.}
\label{reptable}
\end{table}

As a check of our result, let us compute the numbers of states contributing to $\chi^{\text{SO(9),B}}_n$
and $\chi^{\text{SO(9),F}}_n$. Those can be counted by reading each column of Table \ref{reptable},
or by setting $x_i=0$ in \eqref{so9n1} and \eqref{so9n2}.
We see that these two ways give the same values and the numbers of boson states and fermion states are equal:
\bea
\chi^{\text{SO(9),B}}_0,~ \chi^{\text{SO(9),F}}_0 & \rightarrow & 183040~\text{states}, \\
\chi^{\text{SO(9),B}}_1,~ \chi^{\text{SO(9),F}}_1 & \rightarrow & 439296~\text{states}, \\
\chi^{\text{SO(9),B}}_2,~ \chi^{\text{SO(9),F}}_2 & \rightarrow & 465920~\text{states}, \\
\chi^{\text{SO(9),B}}_3,~ \chi^{\text{SO(9),F}}_3 & \rightarrow & 326144~\text{states}, \\
\chi^{\text{SO(9),B}}_4,~ \chi^{\text{SO(9),F}}_4 & \rightarrow & 161280~\text{states}, \\
\chi^{\text{SO(9),B}}_5,~ \chi^{\text{SO(9),F}}_5 & \rightarrow & 56320~\text{states}, \\
\chi^{\text{SO(9),B}}_6,~ \chi^{\text{SO(9),F}}_6 & \rightarrow & 13312~\text{states} , \\
\chi^{\text{SO(9),B}}_7,~ \chi^{\text{SO(9),F}}_7 & \rightarrow & 1920~\text{states}, \\
\chi^{\text{SO(9),B}}_8,~ \chi^{\text{SO(9),F}}_8 & \rightarrow & 128~\text{states}.
\eea
Then we can confirm that the total number of states is equal to $2^{24}$:
\bea
2^{24} = 16777216 & = &
(183040\times 2)\times 1+(439296\times 2)\times 3+(465920\times 2)\times 5
\nn & &
+(326144\times 2)\times 7+(161280\times 2)\times 9+(56320\times 2)\times 11
\nn & &
+(13312\times 2)\times 13+(1920\times 2)\times 15+(128\times 2)\times 17.
\eea
From the first row of Table \ref{reptable}, we see that SO(9) singlet states are decomposed into one
singlet and one 13-dimensional representation of SU(2). This is consistent with the result of
\cite{hlt08,ht10,m10},
and gives another proof of the uniqueness of SO(9)$\times$SU(2) singlet.
The second row of Table \ref{reptable} tells us that SO(9) vector states are decomposed into one 3-dimensional, 
one 7-dimensional, one 11-dimensional and one 15-dimensional representation, which is consistent with
the result of \cite{m10}. 
From the fifth row we see that there is no gauge invariant SO(9) spinor, which means
that the condition of full supersymmetry for the linear term in $X_i^a$ in the expansion of zero 
energy wavefunction is always satisfied, because the condition is in the form that a gauge invariant SO(9) spinor
made of the linear term is equal to zero\cite{m10}.

\section{Discussion}

We have computed the SO(9)$\times$SU(2) character for coordinate independent states
in SU(2) Matrix theory and have decomposed it into
the sum of products of SO(9) and SU(2) characters. It immediately gives the decomposition of
those states into SO(9)$\times$SU(2) representations, and
gives another proof of the uniqueness of the coordinate independent SO(9)$\times$SU(2) singlet state.

A next natural question is if similar calculation can be done in the case of SU($N$) gauge group\cite{micwip}.
Especially it is an interesting question if there are two or more SO(9)$\times$SU($N$) singlet states,
or it is unique as in the SU(2) case.

Another question is if all the states can be constructed by acting $\theta_\alpha^a$ on the unique 
SO(9)$\times$SU(2) singlet state.
We can give a hint for it if we can count numbers of such states and compare the result with Table \ref{reptable}.

\renewcommand{\theequation}{\Alph{section}.\arabic{equation}}
\appendix
\addcontentsline{toc}{section}{Appendix}
\vs{.5cm}
\noindent
{\Large\bf Appendix}
\section{Frobenius formula}
\label{appa}
\setcounter{equation}{0}

The character $\chi(R)$ of a representation $R$ is given by a trace of a group element $g$ over states in $R$:
$\chi(R)=\text{tr}_R(g)$, and we define $\chi(R^k)$ by $\chi(R^k)=\text{tr}_R(g^k)$.
Then the character of $n$-rank antisymmetric tensor product $\text{Alt}_n(R)$ of a representation $R$ can be
computed by the following Frobenius formula:
\beq
\chi(\text{Alt}_n(R))=\sum_{\stackrel{\mbox{$\scriptstyle \sum_{k=1}^nki_k=n$}}
{\mbox{$\scriptstyle i_k:\;\text{nonnegative integer}$}}
}
(-1)^{n+\sum_{k=1}^ni_k}\prod_{k=1}^n\frac{[\chi(R^k)]^{i_k}}{i_k!\cdot k^{i_k}}.
\eeq
Here we count contributions from boson states and fermions states additively.
If states in representation $R$ are fermionic and we count them with minus sign, then
the sign factor in the above formula must be changed from $(-1)^{n+\sum_{k=1}^ni_k}$
to $(-1)^{\sum_{k=1}^ni_k}$.
Since we need explicit expressions in the text, we show some of them for reader's convenience:
\bea
\chi(\text{Alt}_0(R)) & = & 1, \\
\chi(\text{Alt}_1(R)) & = & \chi_1, \\
\chi(\text{Alt}_2(R)) & = & \frac{1}{2}(\chi_1^2-\chi_2), \\
\chi(\text{Alt}_3(R)) & = & \frac{1}{6}(\chi_1^3-3\chi_1\chi_2+2\chi_3), \\
\chi(\text{Alt}_4(R)) & = & 
\frac{1}{24}(\chi_1^4-6\chi_1^2 \chi_2+3\chi_2^2+8\chi_1\chi_3-6\chi_4),
\\
\chi(\text{Alt}_5(R)) & = & 
\frac{1}{120}(\chi_1^5-10\chi_1^3 \chi_2+15\chi_1\chi_2^2+ 
20\chi_1^2 \chi_3
\nn & &
-20\chi_2\chi_3-30\chi_1\chi_4+24\chi_5),
\\
\chi(\text{Alt}_6(R)) & = & 
\frac{1}{720}(\chi_1^6-15\chi_1^4 \chi_2+45\chi_1^2 \chi_2^2 - 
15\chi_2^3+40\chi_1^3 \chi_3
\nn & &
-120\chi_1\chi_2\chi_3+40\chi_3^2
-90\chi_1^2 \chi_4+90\chi_2\chi_4+144\chi_1\chi_5 -120\chi_6),
\\
\chi(\text{Alt}_7(R)) & = & 
\frac{1}{5040}(\chi_1^7-21\chi_1^5 \chi_2+105\chi_1^3 \chi_2^2 - 
 105\chi_1\chi_2^3+70\chi_1^4 \chi_3
\nn & &
 -420\chi_1^2 \chi_2\chi_3+210\chi_2^2 \chi_3 + 
 280\chi_1\chi_3^2-210\chi_1^3 \chi_4 + 
 630\chi_1\chi_2\chi_4
\nn & &
-420\chi_3 \chi_4 + 
 504\chi_1^2 \chi_5-504\chi_2\chi_5-840\chi_1\chi_6 + 
 720\chi_7),
\\
\chi(\text{Alt}_8(R)) & = & 
\frac{1}{40320}(\chi_1^8-28\chi_1^6 \chi_2+210\chi_1^4 \chi_2^2 - 
 420\chi_1^2 \chi_2^3 +105\chi_2^4
\nn & &
+112\chi_1^5 \chi_3 -1120\chi_1^3 \chi_2\chi_3
+1680\chi_1\chi_2^2 \chi_3 +1120\chi_1^2 \chi_3^2
\nn & &
-1120\chi_2\chi_3^2-420\chi_1^4 \chi_4
+2520\chi_1^2 \chi_2\chi_4 -1260\chi_2^2 \chi_4
\nn & &
-3360\chi_1\chi_3 \chi_4 +1260\chi_4^2+ 
 1344\chi_1^3 \chi_5 -4032\chi_1\chi_2\chi_5
\nn & &
+2688\chi_3 \chi_5 -3360\chi_1^2 \chi_6 +3360\chi_2\chi_6 + 
 5760\chi_1\chi_7-5040\chi_8),
\eea
where $\chi_k=\chi(R^k)$.
\section{SO(9) representations and characters}
\label{appb}
\setcounter{equation}{0}

Representations of SO(9) are uniquely specified by the Dynkin label $[q_1,q_2,q_3.q_4]$,
where $q_1,q_2,q_3.$ and $q_4$ are nonnegative integers.
In the context of Matrix theory, even and odd $q_4$ correspond to bosonic and fermionic states respectively.
The highest weight $\mu$ of the representation $[q_1,q_2,q_3.q_4]$ is 
given by the linear combination of fundamental weights
$\mu_1, \mu_2, \mu_3,$ and $\mu_4$: $\mu=q_1\mu_1+q_2\mu_2+q_3\mu_3+q_4\mu_4$, where
\beq
\mu_1=(1,0,0,0),\quad \mu_2=(1,1,0,0),\quad \mu_3=(1,1,1,0),\quad \mu_4=(1/2,1/2,1/2,1/2).
\eeq
Dimension of $[q_1,q_2,q_3.q_4]$ can be computed by the following expression obtained from
Weyl dimension formula:
\bea
\text{dim}[q_1,q_2,q_3.q_4] & = & \prod_{i=1}^4\left(1+\frac{2(q_i+\dots+q_3)+q_4}{9-2i}\right) \nn
 & & \times\prod_{1\leq i <j\leq 4}\left(1+\frac{q_i+\dots+q_{j-1}+2(q_j+\dots+q_3)+q_4}{9-2i-2j}\right) \nn
 & & \times\prod_{1\leq i <j\leq 4}\left(1+\frac{q_i+\dots+q_{j-1}}{j-i}\right),
\eea
where, for $i=4$, expressions as $q_i+\dots+q_3$ should be ignored.

Since an element $w$ of Weyl group of SO(9) acts on
a weight $\lambda=(\lambda_1,\lambda_2,\lambda_3,\lambda_4)$
as sign flip and permutation $\sigma$ of components:
\beq
w\cdot\lambda=(\pm\lambda_{\sigma(1)},\pm\lambda_{\sigma(2)},\pm\lambda_{\sigma(3)},\pm\lambda_{\sigma(4)}),
\eeq
the character for $[q_1,q_2,q_3.q_4]$, denoted by $\chi_{[q_1,q_2,q_3.q_4]}$, is given
by the following expression obtained from Weyl character formula:
\beq
\chi_{[q_1,q_2,q_3.q_4]}\equiv\text{tr}_{[q_1,q_2,q_3.q_4]}
\big[\exp(iJ_{12}x_1+iJ_{34}x_2+iJ_{56}x_3+iJ_{78}x_4)\big]
=\frac{D_{\rho+\mu}}{D_\rho},
\eeq
where $\rho=\mu_1+\mu_2+\mu_3+\mu_4$ is the Weyl vector, and
\beq
D_\lambda=16\sum_\sigma\text{sgn}(\sigma)\prod_{i=1}^4\sin(\lambda_{\sigma(i)}x_i)
=16\det[\sin(\lambda_jx_i)].
\eeq
Table \ref{rep2} shows some correspondences between Dynkin labels and representations which
we usually construct by taking tensor products of vector and spinor representations,
and characters of some of them are given by
\bea
\chi_{[0000]} & = & 1, \\
\chi_{[1000]} & = & 1+(c_1^2+c_2^2+c_3^2+c_4^2), \\
\chi_{[0010]} & = & 4+3(c_1^2+ c_2^2+ c_3^2+ c_4^2)
+(c_1^2c_2^2+c_1^2c_3^2+c_1^2c_4^2+c_2^2c_3^2+c_2^2c_4^2+c_3^2c_4^2)
\nn & &
+(c_1^2c_2^2c_3^2+c_1^2c_2^2c_4^2+c_1^2c_3^2c_4^2+c_2^2c_3^2c_4^2), \\
\chi_{[0001]} & = & c_1^1c_2^1c_3^1c_4^1, \\
\chi_{[2000]} & = & 4+(c_1^2+c_2^2+c_3^2+c_4^2)
\nn & &
+(c_1^2c_2^2+c_1^2c_3^2+c_1^2c_4^2+c_2^2c_3^2+c_2^2c_4^2+c_3^2c_4^2)
+(c_1^4+c_2^4+c_3^4+c_4^4), \\
\chi_{[1001]} & = & 4c_1^1c_2^1c_3^1c_4^1
+(c_1^3c_2^1c_3^1c_4^1+c_1^1c_2^3c_3^1c_4^1
+c_1^1c_2^1c_3^3c_4^1+c_1^1c_2^1c_3^1c_4^3),
\eea
where $c_i^n=2\cos(nx_i/2)$.
\begin{table}
\begin{center}
\begin{tabular}{|r|r|c|}\hline
Dynkin label & dimension & representation \\ \hline
{}[0,0,0,0] & 1 & singlet \\
{}[1,0,0,0] & 9 & vector \\
{}[0,1,0,0] & 36 & 2-rank antisymmetric \\
{}[0,0,1,0] & 84 & 3-rank antisymmetric \\
{}[0,0,0,1] & 16 & spinor \\
{}[0,0,0,2] & 126 & 4-rank antisymmetric \\
{}[1,0,0,1] & 128 & vector-spinor \\
{}[$n$,0,0,0] & $(2n+7)(n+6)!/(7!\cdot n!)$ & $n$-rank symmetric traceless \\ \hline
\end{tabular}
\caption{Correspondences between Dynkin labels and representations}
\label{rep2}
\end{center}
\end{table}

\newcommand{\J}[4]{{\sl #1} {\bf #2} (#3) #4}
\newcommand{\andJ}[3]{{\bf #1} (#2) #3}
\newcommand{\AP}{Ann.\ Phys.\ (N.Y.)}
\newcommand{\MPL}{Mod.\ Phys.\ Lett.}
\newcommand{\NP}{Nucl.\ Phys.}
\newcommand{\PL}{Phys.\ Lett.}
\newcommand{\PR}{Phys.\ Rev.}
\newcommand{\PRL}{Phys.\ Rev.\ Lett.}
\newcommand{\PTP}{Prog.\ Theor.\ Phys.}
\newcommand{\hepth}[1]{{\tt hep-th/#1}}
\newcommand{\arxivhep}[1]{{\tt arXiv.org:#1 [hep-th]}}

\end{document}